\documentclass[lettersize,journal]{IEEEtran}
\usepackage{amsmath,amsfonts}
\usepackage{algorithmic}
\usepackage{algorithm}

\usepackage{float}
\usepackage{array}
\usepackage[caption=false,font=normalsize,labelfont=sf,textfont=sf]{subfig}
\usepackage{textcomp}
\usepackage{stfloats}
\usepackage{url}
\usepackage{verbatim}
\usepackage{graphicx}
\usepackage{cite}
\usepackage{hyperref} 
\hypersetup{
	hidelinks,
	colorlinks=true,
	linkcolor=blue,
	citecolor=blue
}
\begin{document}

	\title{Model Order Reduction for Large-scale Circuits Using Higher Order Dynamic \\ Mode Decomposition}
	\author{
		Na~Liu,~\IEEEmembership{Senior Member,~IEEE,}
		Chengliang~Dai,
		Qiuyue Wu,
		Qiuqi~Li and \\
		Guoxiong~Cai~\IEEEmembership{Member,~IEEE}
	\thanks{This research is supported by the National Natural Science Foundation of China under Grants 62271429 and 62001408, and supported by National Key Research and Development Program of China under Grant 2020YFA0710100.  (\textit{Corresponding author: Guoxiong Cai.})}
	\thanks{N.~Liu,C.~Dai, Q.~Wu and G.~Cai, are with Institute of Electromagnetics and Acoustics, Xiamen University, and Fujian Provincial Key Laboratory of Electromagnetic Wave Science and Detection Technology, Xiamen University, Xiamen 361005, China (e-mail: 34320231150206@stu.xmu.edu.cn; liuna@stu.xmu.edu.cn; qiuyuewu@xmu.edu.cn; gxcai8303@xmu.edu.cn). 
N.~Liu and C.~Dai contributed equally to this work and should be considered co-first authors.} 
	
	\thanks{Q.~Li is with School of Mathematics, Hunan University, Changsha 450082, China (e-mail: qli28@hnu.edu.cn).}}

\markboth{Journal of \LaTeX\ Class Files,~Vol.~$\times\times\times$, No.~$\times\times\times$, February~2025}%
{Shell \MakeLowercase{\textit{et al.}}: A Sample Article Using IEEEtran.cls for IEEE Journals}
\maketitle
\begin{abstract}
Model order reduction (MOR) has long been a mainstream strategy to accelerate large scale transient circuit simulation. Dynamic Mode Decomposition (DMD) represents a novel data-driven characterization method, extracting dominant dynamical modes directly from time-domain simulation data without requiring explicit system equations. This paper first deduces the DMD algorithm and then proposes high order dynamic mode decomposition (HODMD) incorporating delayed embedding technique, specifically targeting computational efficiency in large-scale circuit simulations. Compared with the DMD method, the HODMD method overcomes the problem that the output signal cannot be reconstructed when the spatial resolution is insufficient. The proposed HODMD algorithm is applicable to general circuits and does not impose any constraints on the topology of the pertinent circuit or type of the components. 
Three representative numerical test cases are presented to systematically validate both the computational efficiency and accuracy of the proposed HODMD method.
\end{abstract}


\section{Introduction}
\label {sec:Introduction} 
\IEEEPARstart{T}{he} continuous advancement of integrated circuit (IC) technology enables transistor densities exceeding billions on chips fabricated at 3-nanometer and below process nodes \cite{ref1}. While this technological leap enables system-on-chip (SoC) designs with unprecedented functionality, it simultaneously imposes significant computational challenges on transient circuit simulation — a critical component of electronic design automation (EDA). Modern post-layout full-chip verification typically involves solving systems of differential-algebraic equations (DAEs) with millions of unknowns, a process that can require weeks of computation time even on high-performance computing (HPC) clusters. Such inefficiency directly conflicts with the industry’s need for rapid design iteration, creating a major bottleneck in EDA development. Consequently, accelerating circuit simulation through improved computational capacity and enhanced functional coverage has become imperative to meet escalating IC design demands.

Model order reduction (MOR) has emerged as an indispensable technique in this context\cite{ref2},\cite{ref3},\cite{ref4}. The core idea of MOR is to construct a reduced-order model that accurately preserves the input–output (I/O) behavior of the original high-dimensional circuit system, thereby enabling efficient transient analysis and other computational tasks. The MOR method can mainly be divided into equation-driven and data-driven. Among these approaches, projection-based MOR methods (belong to equation-driven) \cite{ref5},\cite{ref6},\cite{ref7},\cite{ref8}, particularly Krylov subspace techniques like PRIMA \cite{ref9}, have been widely adopted in circuit model reduction. These methods effectively utilize the matrix sparsity while maintaining the moment matching characteristics to achieve the purpose of accelerating the calculation.

Despite their widespread use, projection-based MOR techniques face several significant challenges. These methods often become inefficient when applied to multi-output circuits, primarily due to the computational burden associated with constructing large projection matrices. For multi-output networks, the dimension of the Krylov subspace grows linearly with the number of ports, potentially resulting in reduced-order models whose sizes exceed practical limits and thereby diminishing the expected simulation speedup. Although various strategies have been proposed to address this issue \cite{ref10}, \cite{ref11}, MOR for multi-output circuits remains an open and critical problem. Moreover, projection-based MOR methods exhibit low data utilization efficiency, as they rely heavily on circuit matrices derived from modified nodal analysis equations, making them inherently equation-driven. Consequently, these methods are inapplicable to circuit systems where explicit equations are difficult or impossible to obtain. More critically, this methodology fails to use hidden dynamic information embedded in transient simulation data, resulting in insufficient extrapolation capability when handling complex excitation patterns.

To address the aforementioned challenges, a new data-driven method based on the dynamic mode decomposition (DMD) \cite{ref12},\cite{ref13},\cite{ref14},\cite{ref15} 
has gained increasing attention in the context of MOR. Rooted in Koopman operator theory \cite{ref16}, \cite{ref17}, DMD is a  data-driven technique initially introduced by Schmid \cite{ref12} for analyzing the dynamics of high-dimensional nonlinear systems. DMD decomposes system trajectories into a set of spatial modes, each associated with a distinct temporal behavior characterized by specific frequencies and exponential growth or decay rates.
By analyzing a limited set of state snapshots over a short time window, DMD is capable of extracting the dominant dynamic features of the system. These spatial-temporal modes can then be employed to and predict the future behavior, enabling accurate long-term simulation based solely on observed data.

Based on the above characteristics, the DMD method has been widely adopted in fields such as fluid dynamics, aerospace, and climate science. However, its application in circuit systems remains limited. 
In circuit systems, it is conventional to monitor only a limited number of nodes of interest as output ports, rather than measuring all node voltages and currents. When the number of output ports is insufficient, the resulting snapshot matrix exhibits inadequate dimensionality, which compromises both spatial and temporal resolution and consequently degrades the accuracy of the DMD reconstructed signal.

To address the limitations of standard DMD, this work employs high-order dynamic mode decomposition (HODMD) \cite{ref18},\cite{ref19},\cite{ref20} which enhances the original method by introducing delayed embedding. By mapping one-dimensional time series or low-dimensional snapshots into a higher-dimensional phase space, HODMD facilitates enhanced characterization of circuit system dynamics. HODMD demonstrates the ability to combine a broader range of spatial and temporal models, which leads to a more accurate representation of circuit system dynamics. To the best of our knowledge, this is the first work to introduce the HODMD method for model order reduction of large-scale circuit systems. Compared with conventional projection-based MOR methods, the proposed HODMD method offers several distinct advantages:

\begin{enumerate}
\item The HODMD method is entirely data-driven. Unlike traditional modeling approaches that rely on detailed circuit or structural parameters, the HODMD method directly utilizes measured or provided data to extract spatiotemporal patterns, enabling data-based analysis without the need for explicit circuit equations.

\item Compared to other data-driven methods like neural networks, the HODMD method eliminates the need for extensive training procedures. The HODMD method operates  with a limited set of observational data, thereby avoiding the large datasets and prolonged training times typically required by neural network methods.

\item The HODMD method offers significant computational efficiency. While traditional methods necessitate repeated updates to projection subspace or complex matrix operations, HODMD accomplishes accelerated computation through direct data fitting, effectively circumventing these computationally demanding procedures. This efficiency also establishes a robust foundation for real-time prediction and fault monitoring in integrated circuits.
\end{enumerate}

The remainder of this paper is organized as follows. Section \ref{sec:Circuit EQUATIONS} reviews the background knowledge of circuit. Section \ref{sec:DMD}-A introduces the principles of the DMD method, while section \ref{sec:DMD}-B provides a detailed discussion of the HODMD method. Section \ref{sec:RESULTS}  and \ref{sec:Conclusion} contain the computational examples and the conclusion.

	\section{CIRCUIT EQUATIONS}
	\label {sec:Circuit EQUATIONS} 
A general linear time-invariant circuit can be described using modified nodal analysis (MNA) \cite{ref21},\cite{ref22} as follows:
\begin{equation}\label{eq:eq1}
	\frac{\mathbf{C}d\mathbf{x}(\mathit{t})}{d\mathit{t}} +\mathbf{G}\mathbf{x}(\mathit{t}) =  \mathbf{B}\mathbf{u}_{\text{in}}(\mathit{t})
	$$
	$$
	\mathbf{y}(\mathit{t}) = \mathbf{A}\mathbf{x}(\mathit{t})
\end{equation}
respectively,
\begin{equation}\label{eq:eq2}
	\mathbf{C} = \begin{bmatrix} \mathbf{Q} & \mathbf{0} \\ \mathbf{0} & \mathbf{H} \end{bmatrix}, \quad
	\mathbf{G} = \begin{bmatrix} \mathbf{N} & \mathbf{E} \\ -\mathbf{E}^T & \mathbf{0} \end{bmatrix}, \quad
	\mathbf{x}(t) = \begin{bmatrix} \mathbf{v} \\ \mathbf{i} \end{bmatrix}
\end{equation}
where $ \mathbf{x}(\mathit{t}) \in \mathbb{R}^n $ is the vector of MNA variables, with $\mathbf{v}$ and $\mathbf{i}$ corresponding to the node voltages and branch currents for voltage sources and inductors, respectively, yielding a total of $n$ variables. $ \mathbf{B} \in \mathbb{R}^{n \times n_{\text{in}}} $ and $ \mathbf{A} \in \mathbb{R}^{p \times n} $ are the input and output matrices. $\mathbf{u}_{\text{in}}(\mathit{t}) \in \mathbb{R}^{ n_{\text{in}}}$ represents the input sources and  $\mathbf{y}(\mathit{t}) \in \mathbb{R}^{ p}$ represents  the circuit output.
Here, $ n $ represents the order of the circuit, while $ n_{\text{in}} $ and $ p $ represent the number of independent variables in the circuit inputs and outputs, respectively.

The matrices $\mathbf{C}, \mathbf{G} \in \mathbb{R}^{n \times n}$ represent the conductance and susceptance matrices, respectively, with the exception that the rows corresponding to the current variables are negated. $\mathbf{N}$, $\mathbf{Q}$, and $\mathbf{H}$ are the matrices containing the stamps for resistors, capacitors, and inductors, respectively. $\mathbf{E}$ consists of ones, minus ones, and zeros, which represents the current variables in KCL equations.

Compared with linear circuit equations, nonlinear circuit equations have an additional nonlinear term $\mathbf{F}(\mathbf{x}(\mathit{t}))$. In general, a nonlinear circuit can be described as
\begin{equation}\label{eq:eq25}
	\frac{\mathbf{C}d\mathbf{x}(\mathit{t})}{d\mathit{t}} +\mathbf{G}\mathbf{x}(\mathit{t}) +\mathbf{F}(\mathbf{x}(\mathit{t})) =  \mathbf{B}\mathbf{u}_{\text{in}}(\mathit{t})
$$
$$
\mathbf{y}(\mathit{t}) = \mathbf{A}\mathbf{x}(\mathit{t})
\end{equation}
where $\mathbf{F}(\mathbf{x}(\mathit{t})) \in \mathbb{R}^{n}$ is a nonlinear vector, representing the nonlinear characteristics of semiconductor devices like diodes and transistors.

	\section{HIGHER ORDER DYNAMIC MODE DECOMPOSITION}
\label {sec:DMD} 

This section provides a brief overview of the fundamental principles of DMD. As a data-driven model order reduction technique, DMD extracts dominant spatiotemporal patterns from high-dimensional circuit systems. By decomposing the system's evolution into coherent modes associated with specific frequencies, DMD enables accurate prediction of future system states.

\subsection{Dynamic Mode Decomposition}

For an electronic circuit system, the continuous-time evolution of nodal voltages and currents can be characterized by the output vector $\mathbf{y}(\mathit{t})$. The discrete snapshot sequence $\{\mathbf{y}_0, \mathbf{y}_1, \mathbf{y}_2, \ldots, \mathbf{y}_q\}$
, where each term is a column vector with $p$ rows, $p$ denotes the number of independent variables in the circuit output, and $q$ denotes the number of data snapshots, constitutes a uniform temporal sampling of the continuous-time circuit evolution $\mathbf{y}(\mathit{t})$. 

Consider a discrete state sequence $\mathbf{y}_m \subset \{\mathbf{y}_0, \mathbf{y}_1, \ldots, \mathbf{y}_q\}$, observed at uniformly sampled time steps. The state transition between consecutive observations can be represented through a Koopman operator $\mathbf{L}$ such that:
\begin{equation}\label{eq:eq4}
	\mathbf{y}_{m+1} = \mathbf{L} \mathbf{y}_m
\end{equation}
where $m$ denotes the discrete time index.

The DMD algorithm approximates the Koopman eigenvalues and eigenvectors by constructing consecutive state matrices  $\mathbf{S_1}$ and $\mathbf{S_2} \in \mathbb{R}^{p \times q}$ from observed data, where
\begin{equation}\label{eq:eq6}
	\begin{split}
	\mathbf{S_1} = [\mathbf{y}_0, \mathbf{y}_1, \ldots, \mathbf{y}_{q-1}],\\
	\mathbf{S_2} = [\mathbf{y}_1, \mathbf{y}_2, \ldots, \mathbf{y}_q].
	\end{split}
\end{equation}

Relying on the linear approximation, equation (\ref{eq:eq4}) can be expressed using the data matrices as $\mathbf{S_2} = \mathbf{L} \mathbf{S_1}.$
The best-fit dynamics matrix $\mathbf{L}$ is obtained via
\begin{equation}\label{eq:eq8}
	\mathbf{L} = \mathbf{S_2} \mathbf{S_1^{\dagger}}.
\end{equation}

The DMD modes and eigenvalues correspond to the eigenvectors and eigenvalues of $\mathbf{L}$. Direct eigenvalue decomposition of $\mathbf{L}$ can be computationally prohibitive. To address this,  Schmid's DMD method \cite{ref12} performs eigenvalue decomposition on a reduced-order approximation of $\mathbf{L}$, enabling efficient recovery of its dominant eigenvalues and eigenvectors.

First, perform singular value decomposition (SVD) \cite{ref23} on matrix $\mathbf{S_1} :$
\begin{equation}\label{eq:eq9}
\mathbf{S_1} = \mathbf{U} \mathbf{\Sigma} \mathbf{V}^\mathsf{T}  
\end{equation}
where $ \mathbf{U}^\mathsf{T} \mathbf{U} = \mathbf{I} $ and $ \mathbf{V}^\mathsf{T} \mathbf{V} = \mathbf{I} $, and $ ^\mathsf{T}$ refers to the matrix transpose. $ \mathbf{U} $ and $ \mathbf{V} $ are unitary matrices. Based on the changes of SVD, (\ref{eq:eq8}) becomes
\begin{equation}\label{eq:eq10}
	\mathbf{L} = \mathbf{S_2} \mathbf{V} \mathbf{\Sigma}^{-1} \mathbf{U}^\mathsf{T}. 
\end{equation}

For the data matrix $\mathbf{S_1}$, its theoretical maximum rank is given by $k = \min \{ p, q \}$. However, in many large-scale problems, the important dynamic modes contained in $\mathbf{S_1}$ may have a low rank $ r \ll k $. That is, the singular values on the diagonal of $ \mathbf{\Sigma} $ decay rapidly. In such case, dominant dynamic modes can be extracted by projecting the Koopman operator $\mathbf{L}$ into the subspace spanned by the leading right singular vectors of $\mathbf{L}$. The rank $r$ of the truncated SVD is typically chosen to capture more than 99.9\% of the total energy of the singular values \cite{brunton2022data},
\begin{equation}\label{eq:eqsvd}
	\frac{\sum\limits_{i=1}^{r} \sigma_{i}}{\sum\limits_{i=1}^{k} \sigma_{i}} \geq 99.9\%.
\end{equation}

Define
\begin{equation}\label{eq:eq11}
	\begin{split}
	\mathbf{\tilde{U}} = \mathbf{U}(:, 1:r), \\
	\mathbf{\tilde{\Sigma}} = \mathbf{\Sigma}(1:r, 1:r), \\
	\mathbf{\tilde{V}} = \mathbf{V}(:, 1:r),
	\end{split}
\end{equation}
then $\mathbf{U} \mathbf{\Sigma} \mathbf{V^\mathsf{T}}$ projects $\mathbf{S_1}$ onto an $r$-dimensional subspace. Substituting $\mathbf{S_1} \approx \mathbf{\tilde{U}} \mathbf{\tilde{\Sigma}} \mathbf{\tilde{V}^{\mathsf{T}}}$ into (\ref{eq:eq10}) yields a rank-$r$ estimation of $\mathbf{L}$,
\begin{equation}\label{eq:eq12}
\widetilde{\mathbf{L}} = \widetilde{\mathbf{U}}^{\mathsf{T}} \mathbf{S}_{2} \widetilde{\mathbf{V}} \widetilde{\boldsymbol{\Sigma}}^{-1} \widetilde{\mathbf{U}}^{\mathsf{T}} \widetilde{\mathbf{U}} = \widetilde{\mathbf{U}}^{\mathsf{T}} \mathbf{S}_{2} \widetilde{\mathbf{V}} \widetilde{\boldsymbol{\Sigma}}^{-1}.
\end{equation}

To propagate the original  circuit system, the next key step is computing the eigenvalues and eigenvectors of $ \mathbf{\tilde{L}} $,
\begin{equation}\label{eq:eq13}
	\mathbf{\tilde{L}} \mathbf{W} = \mathbf{W} \mathbf{\Lambda}
\end{equation}
where
\begin{equation}\label{eq:eq14}
	\mathbf{\Lambda} = \begin{bmatrix}
		\lambda_1 & & \\
		& \ddots & \\
		& & \lambda_r
	\end{bmatrix}
\end{equation}
The diagonal matrix $\mathbf{\Lambda}$ contains the eigenvalues, while the columns of $\mathbf{W}$ represent their corresponding eigenvectors. To project these spectral modes back to the original state space, apply the transformation
\begin{equation}\label{eq:eq15}
	\mathbf{\Phi} = \mathbf{S}_2 \mathbf{{\tilde{V}}} \mathbf{{\tilde{\Sigma}}}^{-1} \mathbf{W}.
\end{equation}

The columns of $\mathbf{\Phi}$ are called the DMD modes. Denote
\begin{equation}\label{eq:eq16}
	\mathbf{\Omega} = \frac{\ln \mathbf{\Lambda}}{\Delta t}
\end{equation}
$\mathbf{\Omega}$ contains the growth rate and frequency information of the initial signal and can be used to extrapolate the signal waveform. 

The DMD mode amplitudes are determined by the components of vector $\mathbf{b}$, which depend on the initial measurement vector,
\begin{equation}\label{eq:eq17}
	\mathbf{b} = \mathbf{\Phi}^{\dagger} \mathbf{y}_0.
\end{equation}

Following the derivation of (\ref{eq:eq15}),(\ref{eq:eq16}),(\ref{eq:eq17}), the time-domain representation of $\mathbf{y}(t)$ can be expressed as
\begin{equation}\label{eq:eq5}
	\mathbf{y}(t) = \mathbf{\Phi} \exp(\mathbf{\Omega} t) \mathbf{b}
\end{equation}
The signal $\mathbf{y}(t)$ can be reconstructed through the method proposed above, thereby enabling the evaluation of the continuous states of the current and future circuit systems.

The primary computational cost of the DMD procedure lies in the SVD, which scales as $O(\min(p^{2}q,\, pq^{2}))$ \cite{ref24}. As a data-driven method, DMD extracts dynamical patterns directly from snapshot data without requiring prior knowledge of the circuit governing equations. By analyzing a few initial time steps, DMD projects the circuit system dynamics onto a dominant $r$-dimensional subspace, enabling efficient state prediction while avoiding full state-space computations.

\subsection{Higher Order Dynamic Mode Decomposition}
\begin{figure}[h]
	\centering{\includegraphics[width=0.45\columnwidth,draft=false]{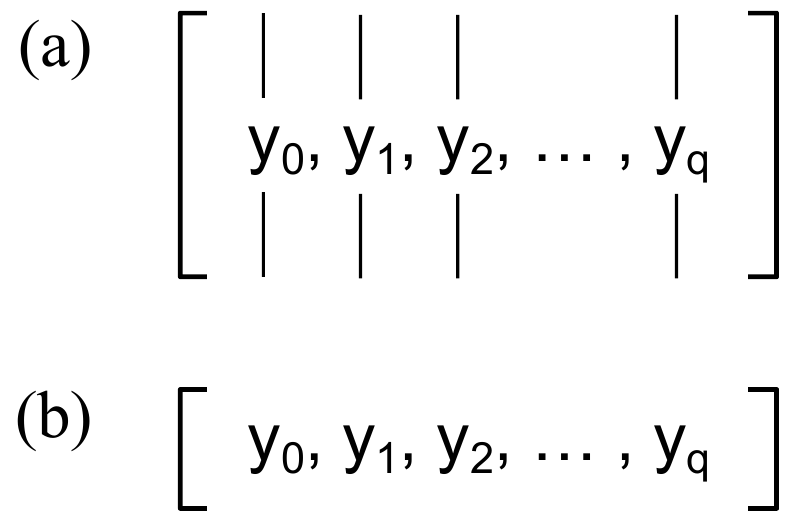}}
	\caption{Snapshot sequence of the DMD method. (a) For general dynamics problem. (b) For circuits analysis} \label{fig:0}
\end{figure}

For the DMD algorithm proposed in the previous section, the number of spatial and temporal modes $r$  in (\ref{eq:eq11}) is determined by the projected Koopman operator $\mathbf{L}$, which maps $\mathbf{S_1}$ to $\mathbf{S_2}$. However, in circuit analysis applications, the DMD method often encounters an under-dimensioned matrix $\mathbf{L}$, which restricts the selectable rank $r$ and consequently fails to fully capture all dynamic modes in the circuit signal $\mathbf{y}(t)$ \cite{yin2023analyzing}.
As illustrated in Fig. \ref{fig:0}, general dynamic systems inherently achieve high spatial dimensionality (the row dimension of the snapshot sequence) through spatial discretization points in the snapshot matrix. In contrast, circuit analysis snapshot sequences are constrained by the limited number of output ports (potentially just one), resulting in severely reduced spatial dimensionality and making it difficult to capture the modes of the system.

To address the limitation of insufficient dimensionality in $\mathbf{L}$, the HODMD method incorporates time-delay embedding \cite{packard1980geometry},\cite{pan2020structure},\cite{takens2006detecting} into the conventional DMD framework, thereby augmenting the expressive capacity of the Koopman operator while maintaining the original spatial resolution. For a single measurement data, the HODMD algorithm employs shift-and-stack operation combined with time-delay embedding techniques to construct augmented data matrix,

\begin{align} \label{19a}
	\mathbf{S}_{\text{aug,1}} &= \begin{bmatrix}
		\mathbf{y}_0 & \mathbf{y}_1 & \cdots & \mathbf{y}_{q-s-1} \\
		\mathbf{y}_1 & \mathbf{y}_2 & \cdots & \mathbf{y}_{q-s} \\
		\vdots & \vdots & \ddots & \vdots \\
		\mathbf{y}_{s-1} & \mathbf{y}_{s} & \cdots & \mathbf{y}_{q-1}
	\end{bmatrix} \tag{19a} \\
	\mathbf{S}_{\text{aug,2}} &= \begin{bmatrix}
		\mathbf{y}_1 & \mathbf{y}_2 & \cdots & \mathbf{y}_{q-s} \\
		\mathbf{y}_2 & \mathbf{y}_3 & \cdots & \mathbf{y}_{q-s+1} \\
		\vdots & \vdots & \ddots & \vdots \\
		\mathbf{y}_{s} & \mathbf{y}_{s+1} & \cdots & \mathbf{y}_q
	\end{bmatrix} \tag{19b}
\end{align}
here, $\mathbf{S}_{\text{aug,1}},\mathbf{S}_{\text{aug,2}} \in \mathbb{R}^{(s \times p) \times (q - s)}$, and $s$ represents the time-delay embedding dimension. 

For the data matrices $ \mathbf{S}_{\text{1}} $ and $ \mathbf{S}_{\text{2}} $ in  (\ref{eq:eq6}) of the traditional DMD method, the total number of singular values of the dynamic mode decomposition is limited by $p$, which represents the number of independent variables. 
When the dimension of $p$ is insufficient, the snapshot matrix may fail to adequately characterize the temporal dynamics of circuit systems. To address this limitation, the HODMD algorithm incorporates $ s - 1 $ time-delay embeddings into the snapshot matrix, extending the snapshot matrix row dimension to $ s \times p $. This augmentation enhances the representation of dynamic modes in the circuit system.
The approximation accuracy of the HODMD method exhibits a positive correlation with the embedding parameter $ s $. For optimal parameter selection, empirical studies \cite{ref29} recommend choosing $ s $ such that the condition $s \times p > 2q$ is satisfied, ensuring sufficient dimensionality for dynamic mode extraction.

The HODMD method proposed in this paper modifies only the construction of the data matrix $\mathbf{S}$, while preserving the subsequent steps of the conventional DMD algorithm. The primary steps of HODMD are illustrated in the pseudocode presented in the following content.

\begin{algorithm}[H]

	\caption{HODMD method}
	\label{alg:nonlinear_connectivity_matrix}
	\textbf{Input:} Given time domain snapshot data, the truncated rank $r$ and the time-delay embedding numbers $s$ \\
	\textbf{Output:} HODMD solution x(t)

	\begin{algorithmic}[1]
		\STATE Obtain $\mathbf{S}_{\text{aug,1}}$ and $\mathbf{S}_{\text{aug,2}}$ from the data set;
		\STATE Perform SVD of $\mathbf{S}_{\text{aug,1}}$ as $\mathbf{S}_{\text{aug,1}} = \mathbf{U} \mathbf{\Sigma} \mathbf{V}^\mathsf{T}  $;
		\STATE Define a low-dimensional projection of $\mathbf{L}$: $\widetilde{\mathbf{L}} = \widetilde{\mathbf{U}}^{\mathsf{T}} \mathbf{S}_{2} \widetilde{\mathbf{V}} \mathbf{{\tilde{\Sigma}}}^{-1} $;
		\STATE Compute the eigenvalues and eigenvectors of $\tilde{\mathbf{L}}$: $\tilde{\mathbf{L}} \mathbf{W} = \mathbf{W} \mathbf{\Lambda}$;
		\STATE Set the modes of $\mathbf{L}$ as $	\mathbf{\Phi} = \mathbf{S}_{\text{aug,2}} \mathbf{{\tilde{V}}} \mathbf{{\tilde{\Sigma}}}^{-1} \mathbf{W}$;
		\STATE Denote $\mathbf{\Omega} = \frac{\ln \mathbf{\Lambda}}{\Delta t}$. The future state in the observable space is given by $\mathbf{y}(t) = \mathbf{\Phi} \exp(\mathbf{\Omega} t) \mathbf{b}$, where $\mathbf{b} = \mathbf{\Phi}^{\dagger} \mathbf{y}_0$.
	\end{algorithmic}
\end{algorithm}

While traditional DMD effectively captures first-order dynamic modes representing linear state transitions, it may struggle with complex dynamics in low-dimensional circuit systems. In contrast, HODMD demonstrates superior potential by employing time-delay embedding to construct more refined approximations of the Koopman operator. This approach incorporates a broader range of spatial and temporal modes, resulting in more accurate dynamic representations without incurring significant computational overhead. Unlike merely increasing spatial resolution, HODMD provides an efficient alternative for enhancing reconstruction accuracy.

\section{NUMERICAL RESULTS }
	\label {sec:RESULTS} 
In this section, three numerical examples are presented to evaluate the performance of the proposed HODMD method for model order reduction of large-scale circuit systems. 
The computation time and speedup ratio are measured using MATLAB R2023b and HSPICE 2024 software on a 64-bit Windows 10 Pro system equipped with an Intel Core i7-CPU at 3.4 GHz and 32 GB of RAM. 
\begin{figure*}[t]
	\centering{\includegraphics[width=\textwidth,draft=false]{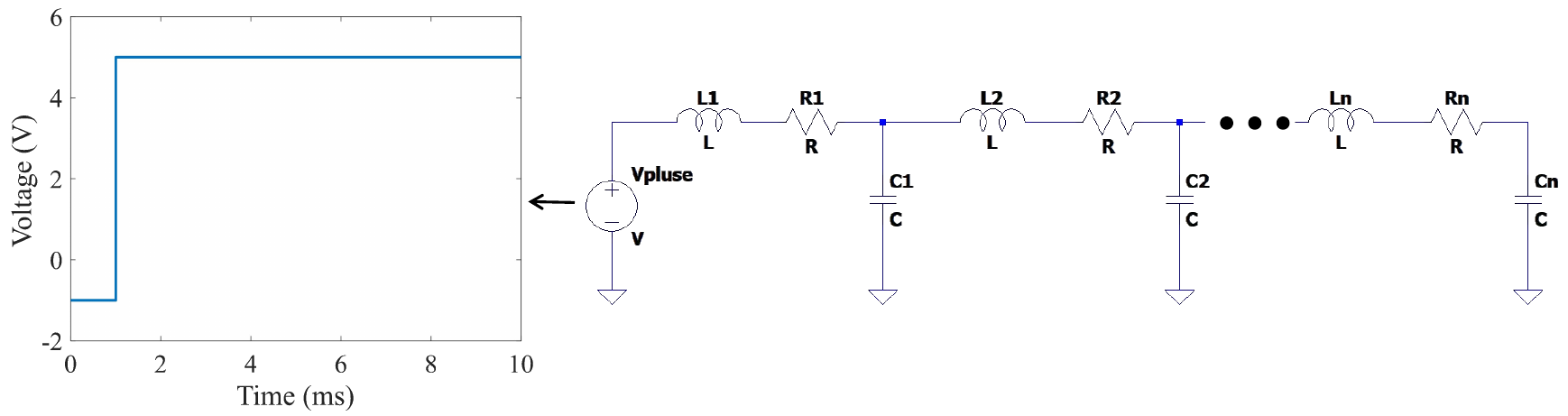}}
	\caption{Cascade linear transmission line circuit with step voltage input (The initial voltage is 0 and the step voltage is 5 V).}
	\label{fig:3}
\end{figure*}

\subsection{Linear Transmission Line Circuit}
In this example, a periodically loaded linear transmission line circuit model is considered, as illustrated in Fig. \ref{fig:3}. This ladder-type transmission line consists of 10,000 identical elementary sections ($n$ = 10,000), each containing series-connected LRC components (one inductor L, one resistor R, and one capacitor C per section). The complete circuit comprises 20,002 nodes, and the load voltage across the terminal capacitor is designated as the output voltage. 

%\begin{figure}[htbp]
%	\centering{\includegraphics[width=1\columnwidth,draft=false]{./picture/4}}
%	\caption{Excitation waveform at the input.} \label{fig:4}
%\end{figure}

\begin{figure}[htbp]
	\centering{\includegraphics[width=1\columnwidth,draft=false]{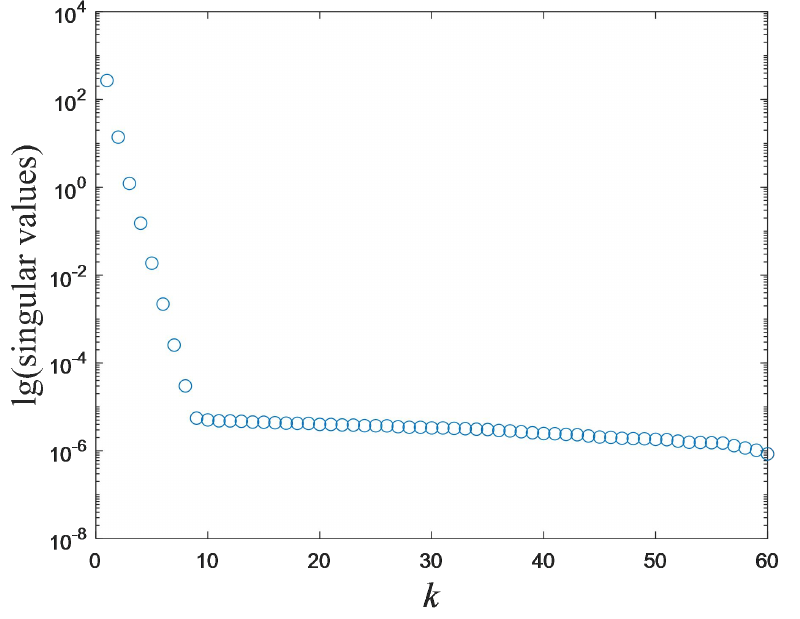}}
	\caption{Singular value distribution of self-built transmission line circuit.} \label{fig:4}
\end{figure}

To generate time-domain snapshots, a step signal with initial voltage of 0 V and step voltage of 5 V (triggered at 1 ms) is applied as the input, as illustrated in Fig. \ref{fig:3}. The simulation spans 2×$10^{-5}$ s with a temporal resolution of 4,000 steps. 
HSPICE simulation results from steps 0 to 400 are used as the training data set for the HODMD method, while subsequent HODMD extrapolation predicts 3,600 time steps of circuit responses. 

\begin{figure}[htbp]
	\centering{\includegraphics[width=1\columnwidth,draft=false]{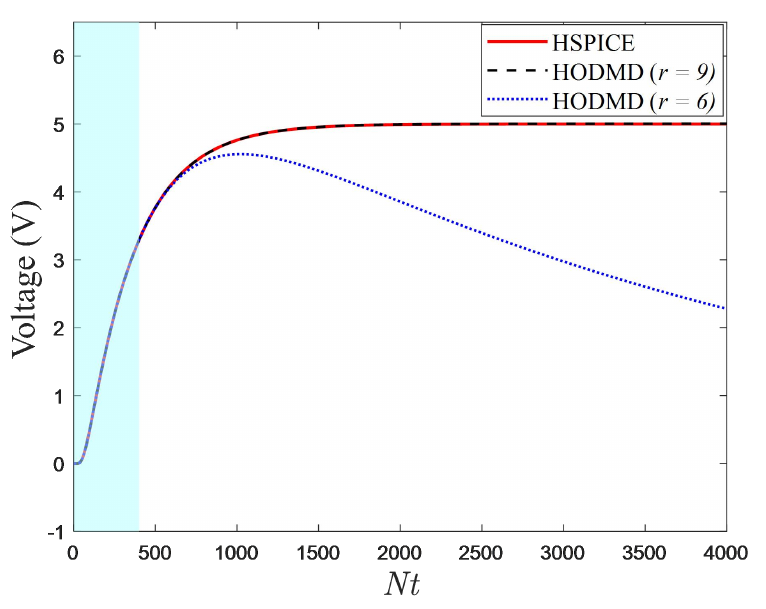}}
	\caption{Comparison of the output response for the transmission line circuit.} \label{fig:5}
\end{figure}
\begin{figure}[htbp]
	\centering{\includegraphics[width=1\columnwidth,draft=false]{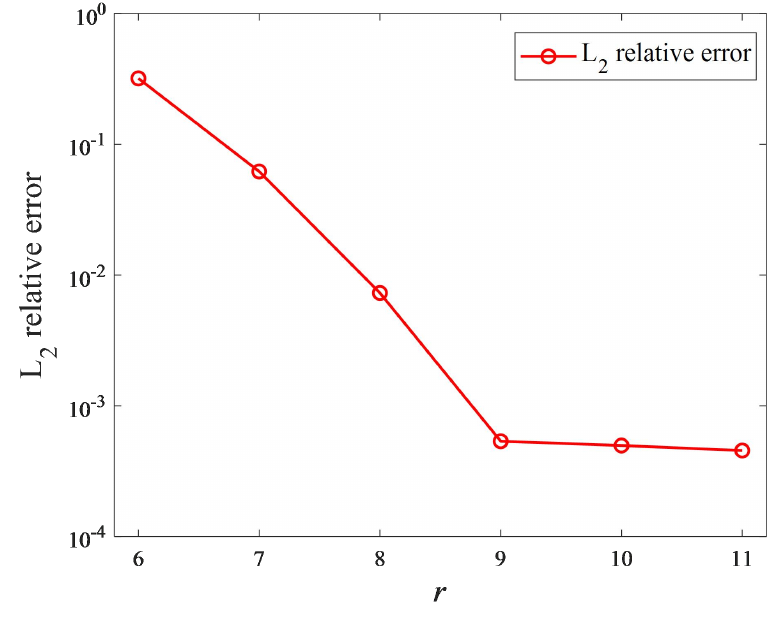}}
	\caption{$L_2$ relative error for different truncated $r$ values of transmission line circuit.} \label{fig:6}
\end{figure}

The singular value distribution of the HODMD method of the linear transmission line model is shown in Fig. \ref{fig:4}. It can be observed that the first nine singular values are orders of magnitudes larger than the the remaining ones. Consequently, by truncating the singular values of the dominant mode \cite{brunton2022data}, the order reduction dimension is determined to be $r$ = 9. Fig. \ref{fig:5} shows a comparison of the output voltage between the extrapolated response using HODMD and the original HSPICE response. The shaded data in Fig. \ref{fig:5} is the snapshot data inputing of the proposed HODMD method. To ensure adequate extraction of spatial and temporal modes, the HODMD method expands the matrix rows by setting parameter $s$ to 60 in (\ref{19a}). The results indicate that when the truncation value $r$ = 9 is selected to retain the dominant modes, the simulation accuracy of the HODMD algorithm is basically the same as that of the HSPICE simulation. The $L_2$ relative error measurement between these two responses is 0.053\%. However, when the number of retained singular values is insufficient, for example, when $r$ = 6 is selected, the results of the HODMD method deviate significantly from those of HSPICE.

Fig.~\ref{fig:6} illustrates the $L_2$ relative error of the signals reconstructed by HODMD for different truncation values of $r$. As $r$ increases, the $L_2$ relative error gradually decreases, indicating improved reconstruction accuracy. However, a larger $r$ also leads to increased matrix dimensions in the computation, resulting in higher computational cost. Therefore, to balance accuracy and computation cost, each model retains the number of singular values corresponding to the dominant modes.

\begin{table}[htbp]
	\centering % 让表格居中
	\caption{Cpu Time And Speedup Of Hspice And Hodmd For \\Linear Transmission Line Circuit}
	\setlength{\tabcolsep}{5mm}
	\renewcommand{\arraystretch}{1.5}  % 1.5倍行距核心设置
	{
		\begin{tabular}{ccc}
			\hline\hline
			Methods       & CPU time (s) & Speed up   \\
			\hline
			HSPICE         	 & 35.14   &     1      \\
			HODMD ($r$ = 9) 	& 4.305   &    8.17  \\
			\hline\hline
	\end{tabular}}
	\label{example2_time}
\end{table}

It is notable that, compared with the HSPICE calculation of the full-order model, the proposed HODMD method provides significant computational acceleration. Table  \ref{example2_time} summarizes a summary of the CPU times for both HSPICE simulations and HODMD calculation. The HODMD method achieves a speedup factor of 8.17×. The total simulation time for HODMD, which includes both the training and extrapolation phases, consists of 4.30 s offline for obtaining the offline snapshot data and 0.005 s online for executing the HODMD method.

\subsection{IBM Power Grid}
In this example, the transient IBM Power Grid (IBMPG) benchmarks \cite{nassif2008power} are investigated with the specifications in Table \ref{example5_time}. IBMPG is a standardized benchmark designed for algorithm performance evaluation in power distribution grid analysis. It employs an equivalent circuit network model to simulate grid characteristics, enabling comparative assessment of algorithmic efficiency and accuracy.

\begin{table}[htbp]
	\centering % 让表格居中
	\caption{Ibm Power Grid Specification}
	\setlength{\tabcolsep}{1.5mm}
	\renewcommand{\arraystretch}{1.5}  % 1.5倍行距核心设置
	{
		\begin{tabular}{ccccccc}
			\hline\hline
			Case       & Node     &  \#R    & \#C & \#L & \#VSource  & \#ISource \\
			\hline
			IBMPG1t    &39,681   &  40,835   & 10,774  & 277 & 14,308 & 10,774  \\
			IBMPG6t    &2,367,183    &  2,450,911 &  761,484 &  381 &  836,239  &  761,484 \\
			\hline\hline
	\end{tabular}}
	\label{example5_time}
\end{table}

\begin{figure}[h]
	\centering{\includegraphics[width=1\columnwidth,draft=false]{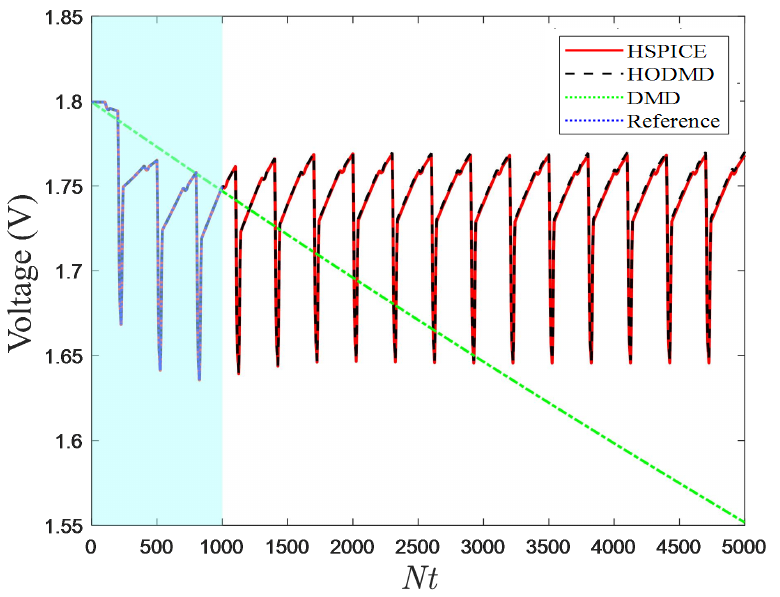}}
	\caption{Comparison of the output response for the IBMPG1t model.} \label{fig:9}
\end{figure}

\begin{figure*}[t]
	\centerline{\includegraphics[width=\textwidth,draft=false]{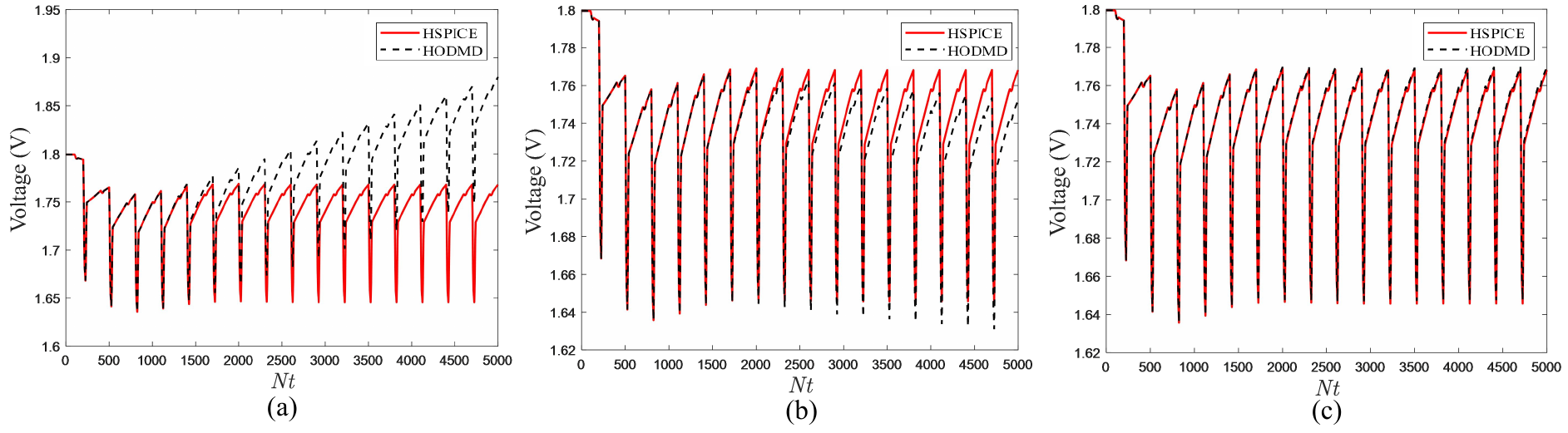}}
	\caption{Output voltage (IBMPG1t) comparison of different augmented matrix rows s : (a) s = 400; (b) s = 450; (c) s = 500.}
	\label{fig:10}
\end{figure*}

Firstly, the effectiveness of the HODMD method is verified through the IBMPG1t, and the voltage signal at node n1-9333-17927 of the IBMPG1t circuit is selected as the output.
The IBMPG1t model consists of 39,681 nodes, 40,835 resistors, 10,774 capacitors, 277 inductors, 14,308 voltage sources and 10,774 current sources. 
The IBMPG1t benchmark provides reference solutions with simulation durations ranging from 0 to 1×$10^{-8}$ s, with a simulation time step of 1×$10^{-11}$ s. The $L_2$ relative error between HSPICE and the reference solution is 0.0013\%, which demonstrates the accuracy of the HSPICE simulation.
The HSPICE simulation consisted of 5,000 time steps, with the HSPICE computed results from Step 1 to Step 1,000 serving as the training data for both DMD and HODMD methods. Both methods extrapolated the subsequent 4,000 time steps. The simulation time for each step is set to 1×$10^{-11}$ s. According to the singular value truncation principle in equation (\ref{eq:eqsvd}), the truncation value $r$ is set to 228. To ensure adequate spatial and temporal resolution, the HODMD method increases the number of matrix rows, with the time-delay embedding parameter $s$ in equation (\ref{19a}) is chosen as 500. Fig. \ref{fig:9} compares the voltage output of IBMPG1t model between the extrapolation circuit signal from HODMD and the original HSPICE response signal. The simulation results clearly demonstrate that the conventional DMD algorithm exhibits significant deviations from the HSPICE computed response, whereas the HODMD method achieves excellent agreement with the HSPICE. The maximum amplitude difference between the HODMD reconstructed response and the original model remained below 0.001 V, with a $L_2$ relative error of 0.061\%.

\iffalse
\begin{figure}[h]
	\centering{\includegraphics[width=1\columnwidth,draft=false]{./picture/8}}
	\caption{Singular value distribution of IBMPG1t model.} \label{fig:8}
\end{figure}
\fi

\begin{table}[htbp]
	\centering % 让表格居中
	\caption{Cpu Time And Speedup Of Hspice And Hodmd \\For Ibm Power Grid}
	\setlength{\tabcolsep}{3mm}
	\renewcommand{\arraystretch}{1.5}  % 1.5倍行距核心设置
	{
		\begin{tabular}{cccc}
			\hline\hline
			Case       & HSPICE time (s) & HODMD time (s) & Speed up   \\
			\hline
			IBMPG1t          & 177.96   &     36.97   &     4.81    \\
			IBMPG6t 		 & 16,058   &   3,358.09  &     4.78  \\
			\hline\hline
	\end{tabular}}
	\label{example3_time}
\end{table}

The total simulation time for the HODMD method, including both training and extrapolation phases, consists of 37.03 s for obtaining the time-domain data and 0.1 s for executing the HODMD method. The CPU time for HSPICE simulations and HODMD predictions is presented in Table \ref{example3_time}, demonstrating that the proposed HODMD method achieves a speedup factor of 4.81× compared to HSPICE simulations.

Fig. \ref{fig:10} compares the HODMD reconstructed waveform of the IBMPG1t model under different augmented matrix row configurations. The results indicate that increasing $s$ within a reasonable range reduces the $L_2$ relative error of the reconstructed signals. The $L_2$ relative errors when the number of matrix rows $s$ = 400, 450 and 500 are 6.29\%, 0.24\% and 0.016\% respectively. When $s$ takes values of  400 and 450, with corresponding $q$ values of 600 and 550 while keeping $p$ fixed at 2 in both cases, the condition $s \times p > 2q $ is not satisfied, resulting in noticeable reconstruction errors. When $s$ takes values of 500, the corresponding values of $p$ is 2 and $q$ is 500. The condition is exactly satisfied, leading to a significantly reduced error as shown in Fig. \ref{fig:10}. Increasing the number of rows in the augmented matrix allows the model to capture more spatial and temporal patterns, thereby enabling a more accurate representation of the system's dynamic behavior.

\begin{figure}[h]
	\centering{\includegraphics[width=1\columnwidth,draft=false]{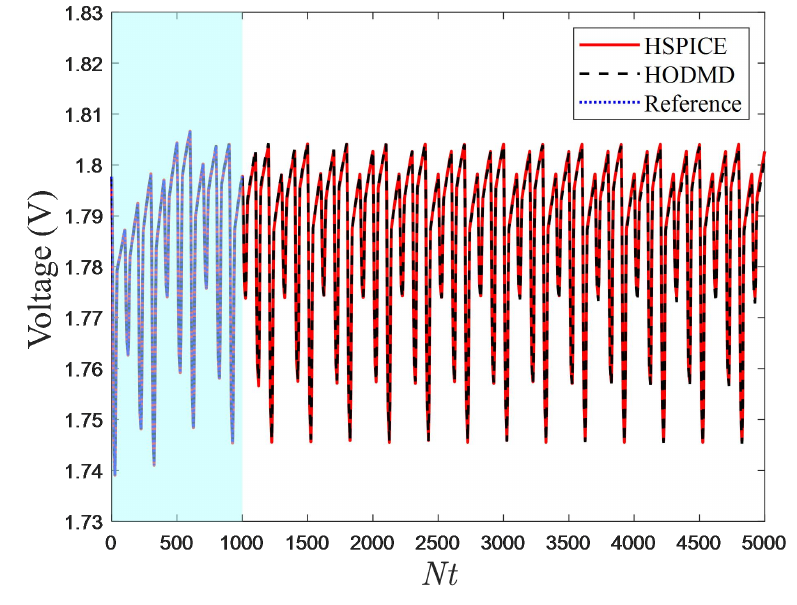}}
	\caption{Comparison of the output response for the IBMPG6t model.} \label{fig:12}
\end{figure}	

\begin{figure*}[t]
	\centering{\includegraphics[width=\textwidth,draft=false]{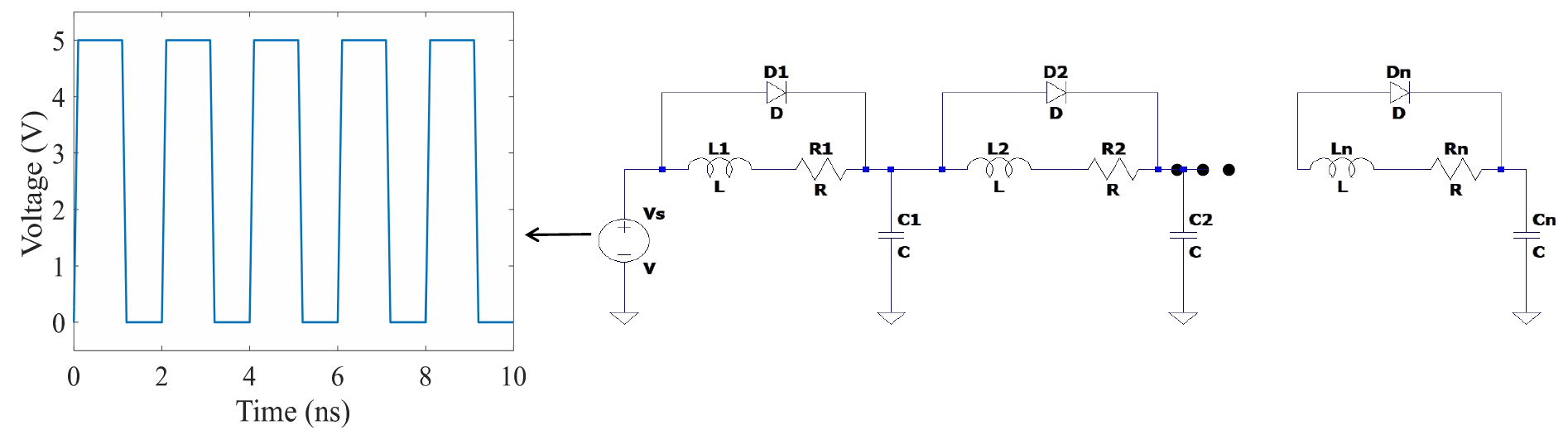}}
	\caption{Nonlinear transmission line circuit with square wave signals as input (the high level is 5 V, the rising and falling edges are both 0.1 ns, and the period is 2 ns).}
	\label{fig:13}
\end{figure*}

Finally, the IBMPG6t model is simulated, and the voltage signal at node n1-1148560-5869360 is selected as the output. The IBMPG6t model consists of 2,367,183 nodes, 2,450,911 resistors, 761,484 capacitors, 381 inductors, 836,239 voltage sources and 761,484 current sources. 
The IBMPG6t benchmark also provides reference solutions with simulation durations ranging from 0 to 1×$10^{-8}$ s, with a time step of 1×$10^{-11}$ s. The $L_2$ relative error between HSPICE and reference solution is 0.014\%. 
To retain the dominant modes, the truncation value $r$ is set to 219 and the matrix rows $s$ is set to 400.  Fig. \ref{fig:12} illustrates the comparison of the time domain waveform for the IBMPG6t model. The simulation time step is set to 1×$10^{-11}$ s, with a total of 5,000 time steps. The time-domain simulation results from HSPICE, spanning from step 0 to step 1,000, are utilized as the snapshot data for the HODMD method. Subsequently, HODMD is employed to extrapolate the remaining 4,000 steps. The results demonstrate that the extrapolated HODMD response exhibits strong consistency with the original HSPICE response, with a maximum amplitude difference of less than 0.001 V and an $L_2$ relative error of 0.028\%. As presented in Table \ref{example3_time}, HODMD method takes 3,358.09 s to reconstruct the signal , including 3,358 s for obtaining the data matrix and 0.09 s for HODMD prediction. The HODMD method achieves a speedup factor of 4.78× compared to traditional HSPICE simulations.

\begin{table}[htbp]
	\centering % 让表格居中
	\caption{$L_2$ Relative Errors For Different Ports Of \\The IBMPG6T Model}
	\setlength{\tabcolsep}{5mm}
	\renewcommand{\arraystretch}{1.5}  % 1.5倍行距核心设置
	{
		\begin{tabular}{ccc}
			\hline\hline
			Port       & $L_2$ Relative Error    \\
			\hline
			
			n0-6110720-141120          & 0.044\%       \\  
		    n1-1799840-10096520          & 0.093\%       \\
		    n3-8123920-6272560          & 0.029\%       \\
			\hline\hline
	\end{tabular}}
	\label{example6_time}
\end{table}

Table \ref{example6_time} shows the $L_2$ relative errors of the other three different output ports, all remaining at a relatively small level.  These results demonstrate that the HODMD method consistently delivers effective acceleration performance, even for circuits containing millions of nodes, thereby offering a promising approach for circuit model order reduction.

\subsection{Nonlinear Transmission Line Circuit}
In this example, different from the previous two linear circuit models, a periodically loaded nonlinear transmission line (NLTL) circuit model based on \cite{nouri2016efficient} illustrated in Fig. \ref{fig:13} is considered. The circuit features a typical cascaded structure composed of 750 identical topological units connected in series (one inductor, one resistor, one capacitor and one diode per section). It contains 1,502 nodes, and the load voltage across the terminal capacitor is designated as the output voltage.

\begin{figure}[htbp]
	\centering{\includegraphics[width=1\columnwidth,draft=false]{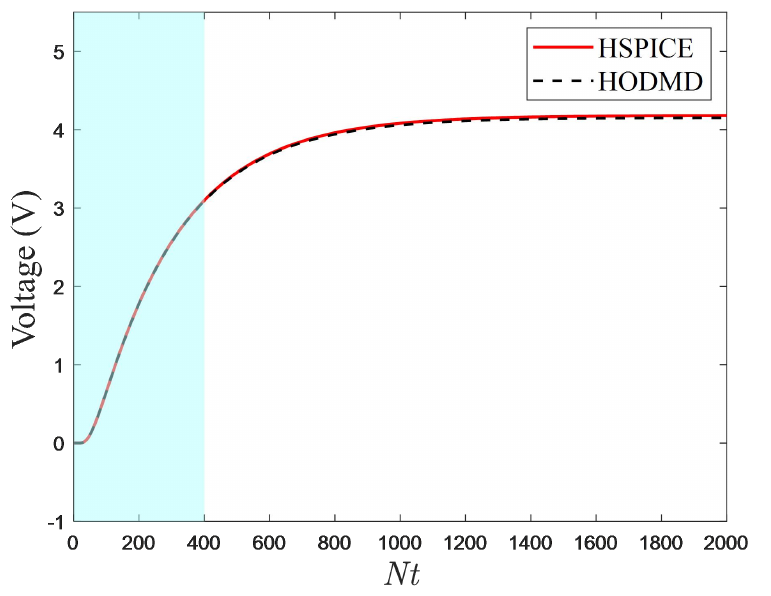}}
	\caption{Comparison of the output response for the NLTL model.} \label{fig:15}
\end{figure}

To generate the time-domain snapshot data, the square wave signal as shown in Fig. \ref{fig:15} is selected as the input. The low level of the square wave signal is 0 V, the high level is 5 V, the duration of the high level is 1ns, the rising and falling edges are both 0.1 ns, and the period is 2 ns. The load voltage across the terminal capacitor of this nonlinear transmission line circuit is designated as the output voltage. The total simulation duration is 1.2×$10^{-4}$ s, and the total number of time steps is 2,000 steps. HSPICE results from steps 0 to 400 are used as the training dataset for the HODMD method, and the remaining 1,600 steps are predicted through HODMD extrapolation.

\iffalse
\begin{figure}[htbp]
	\centering{\includegraphics[width=1\columnwidth,draft=false]{./picture/14}}
	\caption{Singular value distribution of NLTL model.} \label{fig:14}
\end{figure}
\fi

By truncating the singular values of the dominant mode, the order reduction dimension is determined to be $r$ = 25. Additionally, to ensure adequate extraction of spatial and temporal modes, the HODMD method expands the matrix rows by setting parameter $s$ to 60. Fig. \ref{fig:15} shows a comparison of the voltage output between the extrapolation circuit signal using HODMD and the original HSPICE response. The shaded region in Fig. \ref{fig:15} indicates the snapshot data used as input to the proposed HODMD method. The results demonstrate that the HODMD method achieves simulation accuracy comparable to that of full-order HSPICE calculations, while providing substantial computational acceleration.  The measured  $L_2$ relative error between the two reactions is 0.56\%. These results indicate that the HODMD algorithm not only a exhibits strong robustness for linear circuits but also effectively reconstructs signals in nonlinear circuit scenarios.

\begin{table}[htbp]
	\centering % 让表格居中
	\caption{Cpu Time And Speedup Of Hspice And Hodmd For\\Nonlinear Linear Transmission Line Circuit}
	\setlength{\tabcolsep}{5mm}
	\renewcommand{\arraystretch}{1.5}  % 1.5倍行距核心设置
	{
		\begin{tabular}{ccc}
			\hline\hline
			Methods       & CPU time (s) & Speed up   \\
			\hline
			HSPICE          & 2264.56   &     1      \\
			HODMD  & 464.316        &    4.88  \\
			\hline\hline
	\end{tabular}}
	\label{example4_time}
\end{table}
		
Table \ref{example4_time} provides a summary of the CPU times for both HSPICE and the HODMD method. The HODMD method achieves a speedup factor of 4.88×. The total simulation time for HODMD, which includes both the training and extrapolation phases, consists of 464.31 s for obtaining the snapshot data and 0.006 s for executing the HODMD method.

\begin{table}[htbp]
	\centering % 让表格居中
	\caption{ Comparison Of $L_2$ Relative Errors At Output Under Different Input Frequencies Of Square Wave Signals}
	\setlength{\tabcolsep}{8mm}
	\renewcommand{\arraystretch}{1.5}  % 1.5倍行距核心设置
	{
		\begin{tabular}{cc}
			\hline\hline
			Frequency (GHz)       & $L_2$ Relative Error    \\
			\hline
			0.5       	& 0.56\%         \\
			1  			& 0.18\%          \\
			5  			& 0.87\%          \\
			10  		& 0.29\%          \\
			50  		& 0.50\%          \\
			\hline\hline
	\end{tabular}}
	\label{example4_fre}
\end{table}

Table \ref{example4_fre} compares the $L_2$ relative errors at the output port under different input frequencies of square wave signals, and all errors remain below 1\%. These results demonstrate that the HODMD method effectively reconstructs signals in nonlinear circuits, and that its accuracy is robust to variations in input signal frequency.
					
\section{CONCLUSION }
\label {sec:Conclusion}
In this work, an efficient data-driven method is proposed for reduction of large-scale circuit models. Using the proposed HODMD method, the existing snapshot data can be decomposed into temporal patterns and dynamic patterns, thereby achieving the reconstruction and prediction of circuit of circuit signals. In contrast to previously published methods, this method is fully data-driven and does not rely on solving the MNA equations of the circuit, nor is it subject to topological constraints. The simulation results validate the accuracy and robustness of the proposed approach. Furthermore, compared with commercial circuit simulation tools, the method achieves a significant speedup, improving computational efficiency by several times. These findings may inspire new research directions, open up further opportunities in the transient simulation of large-scale circuits and lay the foundation for real-time prediction of circuits.

\bibliographystyle{IEEEtran}
\bibliography{reference.bib}
\newpage

\end{document}